\documentstyle[aaspp4,12pt]{article}
\renewcommand{\ga}{\gamma}

\newcommand{\bea}{\begin{eqnarray}}
\newcommand{\eea}{\end{eqnarray}}
\newcommand{\be}{\begin{equation}}
\newcommand{\ee}{\end{equation}}
\begin{document}

\title{The binary black hole scenario for the BL Lacertae object AO~0235+16}

\author{G.E. Romero $^{1}$, J.H. Fan$^{2}$, S.E. Nuza$^{3}$}


\affil {1. Instituto
 Argentino de Radioastronom\'{\i}a, C.C.5,
(1894) Villa Elisa, Bs. As., Argentina \\2. Center for
Astrophysics, Guangzhou University, Guangzhou 510400, China \\
3. Departamento de F\'{\i}sica J.J. Giambiagi, FCEN, UBA,
Pabell\'on 1, Ciudad Universitaria, 1428 Buenos Aires, Argentina }

\date{\today}

\abstract{Recent analysis of the long term radio lightcurve of the
extremely variable BL Lacertae object AO 0235+16 by Raiteri et al.
(2001) have revealed the presence of recurrent outbursts in this
source with a period of $\sim 5.7\pm 0.5 $ yr. Periodicity
analysis of the optical lightcurve also show evidence for a
shorter period. Here we discuss whether such a behavior can be
explained by a binary black hole model where the accretion disk of
one of the supermassive black holes is precessing due to the tidal
effects of the companion. We estimate the mass of the accreting
hole and analyze under what constraints onto the secondary mass
and the orbital parameters of the system it is possible to provide a viable
interpretation of the available multiwavelength data.
}


\section{Introduction}

The BL Lacertae object AO 0235+16 ($z=0.94$) is one of the most
variable blazars across the entire electromagnetic spectrum. Fan
\& Lin (2000) have compiled the historical optical lightcurves,
which present large amplitude outbursts. At short timescales,
strong radio variability has been found by Quirrenbach et al.
(1992), Romero et al. (1997), and Kraus et al. (1999), among
others. Very rapid (timescales of hours) optical variations were
reported in several opportunities (e.g. Rabbette et al. 1996,
Heidt \& Wagner 1996, Noble \& Miller 1996, Romero et al. 2000a).
The intranight optical duty cycle of this source seems to be close
to 1 (Romero et al. 2002). The X-ray flux also displays
significant and rapid outbursts (e.g. Madejski et al. 1996). At
gamma-ray energies the source has been detected by EGRET
experiment onboard the Compton satellite (Hunter et al. 1993,
Hartman et al. 1999). All this activity makes of AO 0235+16 an
outstanding candidate to probe the most extreme physical
conditions in blazars.

The radio structure of the source has been studied at ground-based
(e.g. Jones et al. 1984, Chu et al. 1996, Chen et al. 1999) and
space (Frey et al. 2000) VLBI resolutions. The object is very
compact at sub-milliarcsecond angular scales. Superluminal
components with velocities up to $\sim30c$ have been detected
(e.g. Fan et al. 1996 and references therein). Chen et al. (1999)
have argued, based on the variations of the position angle of the
superluminal components from 1979 till 1997, that it is possible that the jet is
rotating. The evidence, however, is far from conclusive.

Very recently, Raiteri et al. (2001) have reported the results of
a very extensive multifrequency monitoring of AO 0235+16. On the
long term, variations of 5 magnitudes in the $R$ band and up to a
factor 18 in the radio emission were found. Periodicity analysis
of the radio data based on the discrete autocorrelation function,
the discrete Fourier Transform, and folded lightcurves, covering a
time-span of $\sim25$ yr, reveal the likely existence of a period
of $5.7\pm0.54$ yr (Raiteri et al. 2001). Additional analysis of
the optical lightcurve with the Jurkevich method by Fan et al.
(2002) shows a different periodic signal at $2.95\pm0.15$ yr. The
significance of this latter periodicity is confirmed by Monte
Carlo simulations of random lightcurves (see Fan et al. 2002 for
details).

Periodic signals in the lightcurves of different blazars have been
interpreted by a number of authors in terms of suppermassive
binary black holes systems (e.g. Sillamp\"a\"a et al. 1988, Katz
1997, Villata et al. 1998, Romero et al. 2000b, Rieger \&
Mannhaeim 2000, de Paolis et al. 2002). In the present paper we
will discuss whether the binary black hole hypothesis can be
adapted to the interesting case of AO 0235+16. The existence of
abundant multiwavelength data on this source will
help to constrain the model parameters. In the next section we
shall present the basic features of the precessing jet model for
AGNs. Then we will apply the model to AO 0235+16 and discuss the
results. We close the paper with some brief conclusions.

\section{Supermassive black hole binaries and disk precession}

\begin{figure}
\vbox to7.2in{\rule{0pt}{7.2in}} \includegraphics{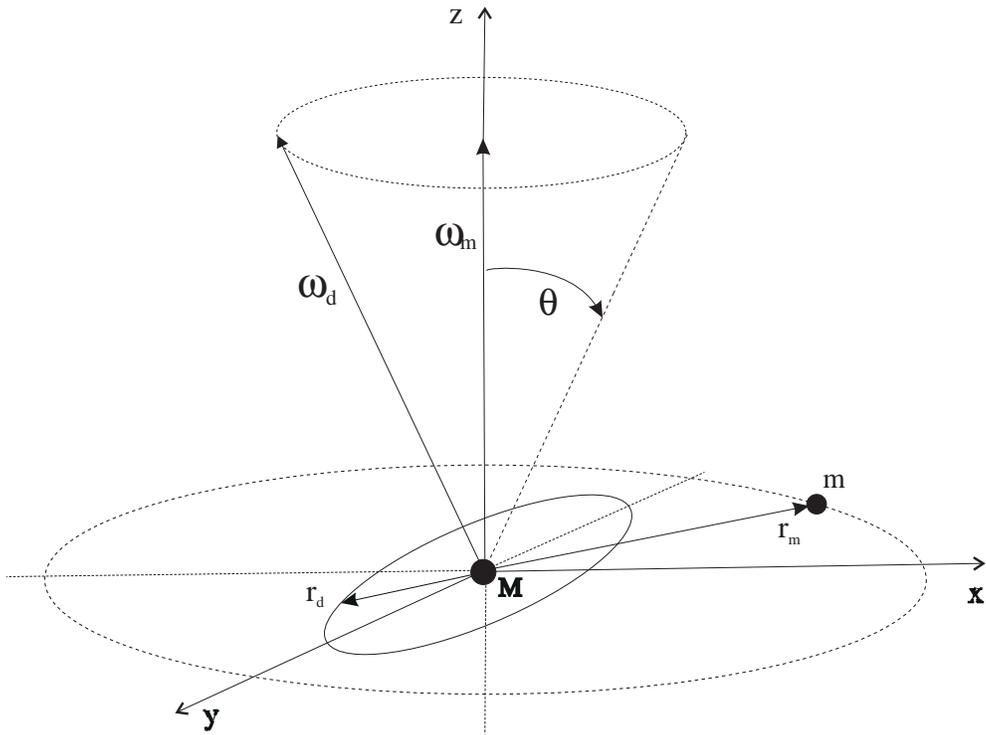} \caption{ A
sketch of a driven precessing accretion disk in a binary black
hole system. Jet direction is indicated by $\vec{ \omega_{\rm
d}}$. Adapted from Romero et al. (2000b) }  \label{f1}
\end{figure}

Supermassive black hole binaries (SBHBs) are the natural result of
galaxy mergers. Their formation and evolution have been
extensively discussed in the literature (e.g. Begelman et al.
1980, Roos 1981, Valtaoja et al. 1989). The fact that many (if not
most) galaxies contain massive black holes and that galaxies often
merge implies a relatively high formation rate of SBHBs. The
current evidence for central engines of active galactic nuclei
formed by massive binary systems includes double nuclei (as in the
case of NGC4486B), wiggly jets (e.g. Kaastra \& Roos 1992), double
emission lines observed in several quasars (Gaskell 1996), and
periodic optical lightcurves as in the case of OJ 287
(Sillanp\"a\"a et al. 1988, Lehto \& Valtonen 1996, Villata et al.
1998).

Geodetic precession of relativistic jets in SBHBs has also been
often discussed in relation to large-scale helical jets (e.g.
Begelman et al. 1980, Roos 1988). This effect is due to the
Lense-Thirring dragging of inertial frames and is much slower than
the tidally-induced precession produced by the gravitational
torque of one of the black holes on the accretion disk of the
other. If we are interested in short timescales we should
focus on the second phenomenon (e.g. Katz 1997, Romero et al.
2000b). Newtonian precession of an accretion disk (which can
be transmitted to the associated jets) has been extensively
studied in the context of galactic binaries and microquasars (e.g.
Katz 1973, 1980; Katz et al. 1982; Larwood 1998; Kaufman Bernad\'o et al.
2002). The general situation in an extragalactic scenario is
depicted in Figure 1. We have two black holes in a close circular
orbit of radius $r_{\rm m}$. One of the holes has an accretion
disk which is non-coplanar with the orbital plane. It is usually
expected that the jet is ejected perpendicularly to the disk
plane, in the direction of the disk angular velocity $\omega_{\rm
d}$. The gravitational torque of the companion black hole onto the
disk will make the innermost part of it to precess, say within a
radius $r_{\rm d}$ where the different parts of the fluid are
efficiently coupled at the sound speed $c_{\rm s}$. This
precession will be transmitted to the jets, which will move with a
precession half-opening angle $\theta$ and a precession velocity
given by (e.g. Katz 1997, Romero et al. 2000b): \be \left|\Omega_{\rm
p}\right|\approx \frac{3}{4} \frac{Gm}{r_{\rm m}^3}
 \frac{1}{\omega_{\rm d}} \cos\theta, \label{prec}
\ee where $G$ is the gravitational constant and $m$ is the mass of
the black hole that exerts the torque upon the disk. By convention
we will call this black hole the ``secondary" and the accreting
hole will be called the ``primary" (its mass will be denoted by
$M$). This does not necessarily imply that $M>m$. The orbital period $T_{\rm
m}$ is related with the black hole masses and size of the orbit by
Kepler's law:

\be r_{\rm m}^3=\frac{G(m+M)T_{\rm m}^2}{4\pi^2}. \label{kepler}
\ee
The ratio between the orbital and the precessing periods can be related through
the disk angular velocity $\omega_{\rm d}=(GM/r_{\rm d}^3)^{1/2}$:
\be
\frac{T_{\rm m}}{T_{\rm p}}=\frac{3}{4}\frac{m}{M} \kappa^{3/2} \left(\frac{M}{m+M}\right)^{1/2} \cos \theta, \label{ratio}
\ee
where $T_{\rm p}=2\pi/\Omega_{\rm p}$ and $\kappa=r_{\rm d}/r_{\rm m}$. Since $\kappa<1$, normally $T_{\rm m}/T_{\rm p}<1$ too. For X-ray binaries in the Galaxy this ratio is typically $\sim 0.1$ (e.g. Larwood 1998).

The precession of the jet results into a variable viewing angle
$\phi=\phi(t)$, which through the flux modulation due to the
Doppler factor $\delta=[\gamma(1-\beta\cos\phi)]^{-1}$ can produce
a periodic signal in the non-thermal jet emission measured in the
observer's frame (e.g. Abraham \& Romero 1999):

\be
S^{\rm obs}(\nu)=\delta^{2+\alpha} S(\nu), \label{S}
\ee
where $\alpha$ is the synchrotron spectral index.

In the next section we will constrain the value of the different parameters for the
application of this scenario to AO 0235+16 and we will then try to
evaluate the likelihood of the binary black hole hypothesis as a
viable explanation of the radio periodicity observed in AO
0235+16.

\section{Models for AO 0235+16}

The non-thermal radio emission of AO 0235+16 is interpreted as incoherent synchrotron
radiation produced by a power-law population of relativistic electrons in the jet of the
object. Hence, the radio periodicity identified by Raiteri et al. (2001) should be related
to processes occurring in the jet. We shall assume that the observed period is the
consequence of the precession of the jet. In the observer's frame, this period is $T^{\rm
obs}_{\rm p}\approx 5.7$ yr. Due to relativistic effects, the intrinsic period in the
blazar will be (e.g., Roland et al. 1994, Rieger \& Mannheim 2000, Britzen et al. 2001):
\be
T^{\rm obs}_{\rm p}=(1+z) \int^{T_{\rm p}}_{0} (1-\beta \cos \phi(t)) dt. \label{Tobs}
\ee
For small viewing angles, as it is the case with AO 0235+16, this yields:
\be
T_{\rm p}\approx \frac{2 \gamma^{2}T^{\rm obs}_{\rm p}}{1+z}. \label{Tp}
\ee

As mentioned by Romero et al. (2000b) for the case of 3C273, a
secondary black hole in a non-coplanar circular orbit around an
accreting black hole must penetrate the outer parts of the disk,
producing optical flares. A similar scenario has been discussed in
connection to OJ 287 by Letho \& Valtonen (1996). Since two
black-hole/disk collisions are expected per orbit, the periodicity
of the optical flares should be $\sim T_{\rm m}/2$.

We shall assume that the optical periodicity found by Fan et al. (2002) in AO 0235+16
corresponds to the disk penetration by the secondary. Correcting by redshift, we get
$T_{\rm m}\approx 3$ yr.  Since this emission is originated in the accretion disk, which
can be considered stationary in the observer's frame, no relativistic corrections should
be applied to this period. Just as an example, we mention that in the case of a jet with a
Lorentz factor $\gamma=2.5$ we have a ratio $T_{\rm m}/T_{\rm p}\sim 0.08$, which is quite
reasonable from a dynamical point of view (Katz 1973, Larwood 1998). Several specific
models will be calculated below.

Estimates of the central black hole mass of AO 0235+16 can be
obtained using high-energy data. The object has been observed by
{\it ROSAT} and {\it ASCA} at soft X-ray energies (Madejski et al.
1996). The {\it ROSAT} data show rapid ($\sim 3$ days) and
significant variability whereas the {\it ASCA} data present a
steady source with a hard power-law energy index $\alpha_{\rm
x}=0.96\pm0.09$ and a flux of 0.3 $\mu$Jy at 1 keV. A small part
of the X-ray emission is expected to be an isotropic field
produced by the innermost part of the accretion disk whereas the
remaining X-rays are probably beamed radiation from the jet, as
suggested by the rapid variability observed by {\it ROSAT}.
Gamma-rays produced close to the black hole will be absorbed in
the isotropic X-ray fields by pair production (Becker \& Kafatos
1995, Blandford \& Levinson 1995). The region at which the opacity
to pair creation drops to 1 for a given energy $E$ defines the
concept of a $\gamma$-sphere: no photons with energy larger than
$E$ will escape from the interior of the corresponding
$\gamma$-sphere. Hence, if we have an adequate model for the
accretion disk generating the absorbing field and an independent
estimate (e.g. through high-energy variability observations) of
the size of a given $\gamma$-sphere, we can calculate the mass of
the black hole (see details in Section 3.3 below). This procedure
has been adopted by a number of
authors in studies of the central objects of gamma-ray blazars
(e.g. Becker \& Kafatos 1995; Fan et al. 1999, 2000; Cheng et al.
1999; Romero et al. 2000c). In particular, Fan et al. (2000) have
estimated the mass of the accreting black hole in AO 0235+16 in
the range $(3.5-5.4)\times10^8$ $M_{\sun}$. In their calculation
they assume a Scwharzschild black hole, with a specific
two-temperature disk model which is responsible for the bulk of
the X-ray emission. In the present work we will relax these
assumptions, calculating a variety of models for both the black
hole and the accretion disk. In addition, we will separate the
X-ray emission in a jet-beamed component and an isotropic
component following the technique introduced by Kembhavi (1993)
and Kembhavi \& Narlikar (1999).

\subsection{Separation of the X-ray components}

In 1987, Browne and Murphy assumed that the X-ray luminosity of
active radio sources, $L_{\rm x}$,  can be separated into two
parts, namely a beamed part, $L_{\rm xb}$ and an unbeamed part,
$L_{\rm xu}$, which gives, $L_{\rm x}=~L_{\rm xb}~+~L_{\rm xu}$.

In standard radio beaming models it is usually assumed that the
ratio of the beamed radio emission, at transverse orientation to
the line of sight, to the extended radio emission is a constant.
Browne and Murphy (1987) extended this to the X-ray luminosity and
assumed that the beamed X-ray luminosity at transverse orientation
is also proportional to the extended radio emission, i.e. $L_{\rm
xb}(90\; \rm{deg})=A\; L_{\rm r,\; ext}$, with $A=$ constant. The
beamed luminosity for an inclination angle $\phi$ between the beam
direction and the line of sight is $L_{\rm xb}(\phi)=g_{\rm
x}(\beta,\phi)L_{\rm xb}(90\;\rm{deg})$, where $g_{\rm
x}(\beta,\phi)$ is the X-ray beaming factor, given by

\be g_{\rm x}(\beta,\phi)={\frac{1}{2}}[(1-\beta
\cos\phi)^{-(2+\alpha_{\rm x})}+(1+\beta
\cos\phi)^{-(2+\alpha_{\rm x})}]. \label{gx} \ee Here, $\beta$ is
the bulk velocity of the beamed flow in units of $c$, and
$\alpha_{\rm x}$ is the spectral index of the beamed X-ray
emission. The expression $\beta \cos\phi$ can be obtained for each
source from the following relation derived by Orr \& Browne (1982) for radio quasars:

\be R_{\rm radio}=R_{90}{\frac{1}{2}}[(1-\beta
\cos\phi)^{-2}+(1+\beta \cos\phi)^{-2}], \label{gr}
 \ee
where $R_{\rm radio}=L_{\rm rc}/L_{\rm r\; ,ext}$ is the
core-dominance ratio with $L_{rc}$ and $L_{r,ext}$ being the core
and extended luminosities, and $R_{90} = 0.024$ according to a
statistical analysis by Orr \& Browne (1982) for a sample of 38
flat-spectrum ($S_\nu \propto \nu^{-\alpha}$, $\alpha=0$) quasars
taken from Jenkins et al. (1977) sample. Then the ratio of the
beamed  to unbeamed X-ray luminosity can be written as:

\be R_{\rm x} = {\frac{L_{\rm xb}}{L_{\rm xu}}} =R_{\rm tx}g_{\rm
x}(\beta,\phi), \;\;\;\; R_{\rm tx}={\frac{L_{\rm
xb}(90\;\rm{deg})}{L_{\rm xb}}}. \label{ratio}
 \ee
In this latter expression $R_{\rm tx}$ is assumed to be a
constant.

There are two problems with the Browne-Murphy model, i.e., a) it
is not suitable for radio quiet quasars, and b) the correlation
$L_{\rm xu} \propto L_{\rm r,\;ext}$, depends on the redshift (see
Kembhavi \& Narlikar, 1999). In order to overcome these problems
Kembhavi (1993) proposed a beaming model that uses the formalism
suggested by Browne and Murphy but with a different scheme for
separating the X-ray luminosity into beamed and isotropic parts.
He considered a subset of 34 quasars with $L_{\rm
rc}/L_{\rm r,\;ext}>10$ (selected from Browne and Murphy) for
which a significant correlation  was found between $\log L_{\rm x}$
and $\log L_{\rm rc}$. Because the fit is dominated by beamed
emission, the $\log L_{\rm x}$ -- $\log L_{\rm rc}$ relation
suggests that there is a relation between the beamed X-ray and
radio components. Following this method, we considered a sample of
19 gamma-ray loud blazars with $L_{\rm rc}/L_{\rm r,\;
ext}>10$ and found a correlation
\be \log \left(\frac{L_{\rm x}}{\rm{W Hz^{-1}}}\right)~=~(0.64\pm0.14)\log \left(\frac{L_{\rm rc}}{\rm{W Hz^{-1}}}\right) + 3.49.\ee
Then we assumed that a relation
\be \log \left(\frac{L_{\rm x}}{\rm{W Hz^{-1}}}\right)~=~0.64\log \left(\frac{L_{\rm rc}}{\rm{W Hz^{-1}}}\right) + \log k \ee
holds for all gamma-ray loud blazars, where $k$ is a constant. In this
 sense, the total X-ray  luminosity is given by
\bea
L_{\rm x} = L_{\rm xb}+L_{\rm xu}&=&
L_{\rm xb}\left(1+{\frac{1}{R_{\rm tx}g_{\rm x}(\beta,\phi)}}\right)\label{totx}\\ 
&=& k\;L_{\rm rc}^{0.64}\left(1+{\frac{1}{R_{\rm tx}g_{\rm x}(\beta,\phi)}}\right).
\label{totx} \eea

The two constants, $\log k = 3.48$ and $R_{\rm tx}=L_{\rm
xb}(90\;\rm{deg})/L_{\rm xu}=5.9\times10^{-3}$, were determined by
minimizing $\Sigma\left[\log (L_{\rm x}/L_{\rm x}^{\rm
obs})\right]^2$ as in Browne \& Murphy (1987) and Kembhavi (1993).
For the specific case of AO 0235+16 we get (all specific luminosities expressed in units of W Hz $^{-1}$):
$\log L_{\rm rc}=27.6$, $\log L_{\rm r,\;ext}=26.41$, $\log L_{\rm x}=21.97$, and $L_{\rm xb}/L_{\rm
xu}=1030$. This means that only a fraction $\sim 10^{-3}$ of the
total X-ray flux can be attributed to the accretion disk. A detail
discussion is presented in a separating paper by Fan, Romero, Lin,
and Zhang (2003, in preparation).

\subsection{Characterization of the models}

In order to make quantitative estimates for the possible binary
systems we shall follow the analytical treatment of Becker \&
Kafatos (1995) to fix the mass of the accreting black hole \footnote{Throughout this paper we assume a Hubble constant $H_0= 100h$ km s$^{-1}$ Mpc$^{-1}$, with $h=0.75$, and a deceleration parameter $q_0=0.5$}. In
particular we shall assume that the inner disk emission can be
represented by models where the intensity has a dependency
$I(E,\;R)\propto E^{-\alpha_{\rm x}}R^{-\xi}$, where $E$ is the
photon energy and $R$ is the radial distance on the disk ($R_{\rm
min}\leq R\leq R_0$). The parameter $\xi$ characterizes the kind of
the disk emission structure. A value $\xi=3$ corresponds to
two-temperature disks where electrons mainly cool through Compton
losses and protons through Coulomb interactions (e.g. Shapiro et
al. 1976). These disks have a flux with a nearly power-law dependence on the
radius $F(R)\propto R^{-3}$ (as the Shakura-Sunyaev disks) and an X-ray power-law
spectrum with an index
\be \alpha_{\rm x}=-\frac{3}{2}+\sqrt{\frac{9}{4}+\frac{4}{y}} \ee
in the source frame, where $y\sim1$ is the Compton $y$ parameter
(e.g. Eliek \& Kafatos 1983).  A value $\xi=0$ corresponds to models of uniform bright
at X-rays (no dependence on $R$) with a single-temperature hot corona that cools by interactions with soft
photons from an underlying cool disk (e.g. Liang 1979). We shall
consider models with Schwarzschild or Kerr black holes, with disks
of single or two temperatures, and different outer radii $R_0$ (the
inner radii $R_{\rm min}$ corresponds to the last stable orbit in each kind of
black hole). We shall name these models, following Romero et al.
(2000b), from model A to H. They are defined in Table
\ref{models}. The radii are expressed in units of the gravitational radius $R_{\rm
g}=GM/c^2$. The extent of the X-ray emitting region is not well known. Typical values for
$R_0$ are $\sim 30\; R_{\rm g}$  (e.g. Shapiro et al. 1976), but higher values are possible. We consider cases with
$R_0=30\; R_{\rm g}$ and $R_0=100\; R_{\rm g}$.

\begin{table}
\caption[]{Models for the accreting black hole in AO 0235+16.}
\label{models}
\begin{tabular}{l l l l}
\noalign{\smallskip} \hline \noalign{\smallskip} Model & BH Type &
Disk Type & $R_{0} /R_{\rm g}$ \cr \noalign{\smallskip} \hline
\noalign{\smallskip} A & Schwarzschild & $\xi = 3$ & 100 \cr B &
Schwarzschild & $\xi = 0$ & 100 \cr C & Schwarzschild & $\xi = 3$
& 30   \cr D & Schwarzschild & $\xi = 0$ & 30 \cr E & Kerr & $\xi
= 3$ & 100 \cr F & Kerr & $\xi = 0$ & 100 \cr G & Kerr & $\xi = 3$
& 30  \cr H & Kerr & $\xi = 0$ & 30
 \cr \noalign{\smallskip} \hline
\end{tabular}
\end{table}

\subsection{Mass of the accreting black hole}

For each model in Table \ref{models} we can now estimate the
optical depth to pair creation for photons of energy $E$
propagating through the accretion disk field:

\be \tau_{\rm x \ga}(E_{\gamma},z)=\int^{\infty}_z
\alpha_{\rm x \ga}(E_{\gamma},z_*)\;dz_*,\label{taudef} \ee
where
$\alpha_{\rm x \ga}$ is the photon-photon absorption coefficient
along the rotation axis of the disk (coincident with the jet
that is nearly pointing to the observer). This coefficient can be
calculated as in Becker \& Kafatos (1995)--see also Cheng et al.
(1999)-- using the photon--photon cross section given by (e.g. Lang 1999):
\be
\sigma(E_{\rm x},\;E_{\gamma})=\frac{\pi r_0^2}{2}(1-\varsigma^2)
\left[2 \varsigma(\varsigma^2-2)+(3-\varsigma^4) \ln \left(\frac{1+\varsigma}{1-\varsigma} \right)\right],\label{sigma}
\ee
where
\be
\varsigma=\left[1-\frac{(m_{\rm e}c^2)^2}{E_{\rm x}E_{\gamma}}\right]^{1/2},
\ee
$r_0=2.818 \times 10^{-13}$ cm is the classical electron radius, $m_{\rm e}$ its mass, and $E_i$ the energy of the
interacting photons.

For calculation purposes, we can write the disk X-ray intensity as:

\be
I(E_{\rm x},\;R)=I_0 \left(\frac{E_{\rm x}}{m_{\rm e}c^2} \right)^{-\alpha_{\rm x}}\left(\frac{R}{R_{\rm g}}\right)^{-\xi}. \label{I}
\ee
Then, the differential absorption coefficient will be
\be
d\alpha_{{\rm x}\gamma}=\frac{I}{c E_{\rm x}} \sigma(E_{\rm x},\;E_{\gamma})(1-\cos\Theta) dE_{\rm x}d\Omega, \label{alpha}
\ee
where $\Theta$ is the angle between $d\Omega$ and the direction of propagation of the $\gamma$-ray. When these rays
propagate mainly along the rotation axis $z$, as expected for jet beamed emission, we can use eq. (\ref{taudef}) and introduce geometrical simplifications due to the axial symmetry. After some algebra, this yields:
\be
\tau_{{\rm x} \gamma} (E_{\gamma},\; z)\cong \frac{AR_{\rm g}}{2\alpha_{\rm x}+3}\left(\frac{z}{R_{\rm g}}\right)^{-2\alpha_{\rm x}-3}\left(\frac{E_{\gamma}}{4m_{\rm e}c^2 }\right)^{\alpha_{\rm x}}.\label{tau}
\ee
Here,
\be
A\equiv\frac{\pi I_0 \sigma_{\rm T} \Psi (\alpha_{\rm x})}{(2\alpha_{\rm x}+4-\xi)c}
\left[\left(\frac{R_{\rm 0}}{R_{\rm g}}\right)^{2\alpha_{\rm x}+4-\xi}-
\left(\frac{R_{\rm min}}{R_{\rm g}}\right)^{2\alpha_{\rm X}+4-\xi}\right],
\label{A} \ee
with $\sigma_{\rm T}$ the Thomson cross section, and
$\Psi(\alpha_{\rm x})$ a function plotted in Fig. 1 of Becker \&
Kafatos' paper ($\Psi(\alpha_{\rm x})\approx 0.18$ for AO 0235+16).
The intensity $I_0$ can be estimated from the
observed X-ray flux, $F_{\rm keV}$, by equation (5.1) of the same
paper along with the condition imposed by our eq. (\ref{totx}).

Since it is an observational fact that AO 0235+16
emits photons with energies $E=1$ GeV (Hartman et al. 1999) we can
impose the condition $\tau_{\rm x \ga}\sim 1$ for photons of such
energy. This condition along with an independent estimate of the
size of the gamma-ray emitting region obtained from high-energy
variability observations allows to estimate the mass of the
accreting black hole through eq. (\ref{tau}). Notice that the mass of the central black hole determines the gravitational radius $R_{\rm g}$. The constraint on the size of the
$\gamma$-spheres is: \be z_{\gamma}\leq c  t_{\rm v}
\frac{\delta}{1+z}\;\;\;\;\rm{cm}.\label{tv} \ee For AO 0235+16
$t_{\rm v}\sim 3$ days (Fan et al. 2000). The Doppler factor
$\delta$ of the underlying jet flow is unknown. It should be
significant smaller than the extreme Doppler factor inferred for
the superluminal component, which are usually interpreted as
relativistic perturbations or shocks propagating downstream. Fan
et al. (2000) estimate a value $\delta\sim 2$. Madejski et al.
(1996) give a higher value $\delta\geq3.1$. Zhang et al. (2002) suggest $\delta=8.9$. We will
perform our calculations for three different values: $\delta=2$,
$\delta=5$, and $\delta=10$.

\begin{table}
\caption[]{Results for different models with $\delta= 2$.}
\begin{tabular}{l c c c c c }
\noalign{\smallskip} \hline \noalign{\smallskip} Model & $M$ & $m^{\rm max}$
& $r^{\rm max}_{\rm m}$ & $r_{\rm d}/r^{\rm max}_{\rm m}$ \cr ($\delta$ $=$ 2) & (10$^{8}$ $M_{\odot}$) &
(10$^{8}$ $M_{\odot}$) & (10$^{16}$ cm)&  \cr \noalign{\smallskip}
\hline \noalign{\smallskip}

A & 6.0 & 320.7 & 10.0 & 0.20 \cr

B & 3.0 & 898.1 & 14.0 & 0.11 \cr

C & 14.1 & 93.4 & 6.9 & 0.43 \cr

D & 9.9 & 153.7 & 7.9 & 0.32 \cr

E & 9.4 & 163.8 & 8.1 & 0.30 \cr

F & 3.0 & 893.6 & 14.0 & 0.11 \cr

G & 23.1 & 50.2 & 6.0 & 0.67 \cr

H & 10.0 & 151.5 & 7.9 & 0.32 \cr \noalign{\smallskip} \hline
\end{tabular}
\label{M1}
\end{table}

\begin{table}
\caption[]{Results for different models with $\delta= 5$.}
\begin{tabular}{l c c c c c }
\noalign{\smallskip} \hline \noalign{\smallskip} Model & $M$ & $m^{\rm max}$
& $r^{\rm max}_{\rm m}$ & $r_{\rm d}/r^{\rm max}_{\rm m}$ \cr ($\delta$ $=$ 5) & (10$^{8}$ $M_{\odot}$) &
(10$^{8}$ $M_{\odot}$) & (10$^{16}$ cm)&  \cr \noalign{\smallskip}
\hline \noalign{\smallskip}

A & 18.7 & 64.7 & 6.3 & 0.16 \cr

B & 9.4 & 164.8 & 8.1 & 0.09 \cr

C & 44.4 & 25.7 & 6.0 & 0.38 \cr

D & 31.0 & 36.4 & 5.9 & 0.26 \cr

E & 30.0 & 38.2 & 5.9 & 0.25 \cr

F & 9.4 & 164.8 & 8.1 & 0.09 \cr

G & 72.7 & 17.1 & 6.5 & 0.65 \cr

H & 31.3 & 36.0 & 5.9 & 0.27 \cr \noalign{\smallskip} \hline
\end{tabular}
\label{M2}
\end{table}

\begin{table}
\caption[]{Results for different models with $\delta= 10$.}
\begin{tabular}{l c c c c c }
\noalign{\smallskip} \hline \noalign{\smallskip} Model & $M$ & $m^{\rm max}$
& $r^{\rm max}_{\rm m}$ & $r_{\rm d}/r^{\rm max}_{\rm m}$ \cr ($\delta$ $=$ 10) & (10$^{8}$ $M_{\odot}$) &
(10$^{8}$ $M_{\odot}$) & (10$^{16}$ cm)&  \cr \noalign{\smallskip}
\hline \noalign{\smallskip}

A & 44.4 & 25.7 & 6.0 & 0.15 \cr

B & 22.3 & 35.8 & 5.6 & 0.09 \cr

C & 106.0 & 12.8 & 7.1 & 0.38 \cr

D & 73.7 & 16.9 & 6.5 & 0.26 \cr

E & 70.5 & 17.5 & 6.4 & 0.25 \cr

F & 22.3 & 35.8 & 5.6 & 0.09 \cr

G & 173.0 & 9.1 & 8.2 & 0.66 \cr

H & 74.5 & 16.7 & 6.5 & 0.28 \cr \noalign{\smallskip} \hline
\end{tabular}
\label{M3}
\end{table}

The results of our estimates of the mass of the accreting black
hole are shown in Tables \ref{M1} - \ref{M3}. The obtained
values for the different models range from $3\times 10^8$
$M_{\sun}$ (e.g. models B nd F for $\delta=2$) up to $\sim17\times10^9$
$M_{\sun}$ ( model G for $\delta=10$). The mass we obtain here for
the case considered by Fan et al. (2000) --model C in Table
\ref{M1}-- is higher because of the refinement
introduced in this paper with the separation of the isotropic and
beamed X-ray components.

We turn now to the secondary black hole mass and the orbital
parameters in order to establish at least an upper bound on them.

\subsection{Mass of the secondary}

A close supermassive black hole system will lose energy by
gravitational radiation and these losses will affect the orbital
parameters. Hence, the errors in the determination of any periodic
signal related to the orbital motion should impose an upper bound
on these losses. From the work by Fan et al. (2002) we find that
in the case of AO 0235+16, $\Delta T_{\rm m}/T_{\rm m}\sim0.05$.
The orbit of the binary will decay on a timescale of (e.g. Shapiro
\& Teukolsky 1983):

\be \tau_0=|r/\dot{r}|
       \sim \frac{5}{256}
              \frac{c^5}{G^3}
              \frac{r_{\rm m}^4}{(m+M)^2}
              \frac{1}{\mu}, \label{tau_0} \ee
where $\mu=mM/(m+M)$. We can rewrite the orbital radius given by eq. (\ref{kepler}) as:
\be
r_{\rm m}\approx 1.4\times 10^{16}\;(m_8+M_8)^{1/3}\;\;\; {\rm cm}. \label{rm}
\ee
Here we have used $T_{\rm m}=3$ yr and the masses are expressed in units of $10^8 M_{\odot}$. With this, eq. (\ref{tau_0}) becomes:
\be
\tau_0\approx 2.8 \times 10^{5} \frac{(m_8 + M_8)^{1/3}}{m_8 M_8}\;\;\;{\rm yr}.
\ee

The relative constancy of the period found by Fan et al. (2002)
implies that $\tau_0>10^3$ yr, which translates into a maximum
possible value of the secondary black hole mass. Eq.
(\ref{rm}) then imposes a maximum value to $r_{\rm m}$. For
calculation purposes we shall adopt $\tau_0=10^3$ yr (in the
source frame) in order to obtain upper limits for parameters.
Shorter timescales are unlikely since the system would be in the
final steps before the merger and other observational consequences
should then manifest (like a tidal disruption of the disk, which
is contradicted by the observation of a persistent jet). The
resulting values for both $m$ and $r_{\rm m}$ are shown in the
third and fourth columns of Tables \ref{M1} -- \ref{M3}.

\section{Analysis of the results}

Uniform disk precession will occur in the scenario here discussed
only if the sound crossing time through the disk is considerably
shorter than the characteristic precession period induced by the
secondary. This allows the bending waves (which propagates at a
velocity $v\sim c_{\rm s}$ through the disk) to efficiently
couple the different parts of the fluid in order to adjust
the precession rate to a constant value (Papaloizou et al. 1998).
This is confirmed by the numerical simulations performed by
Larwood et al. (1996).

\begin{figure}
\vbox to7.2in{\rule{0pt}{7.2in}} \includegraphics{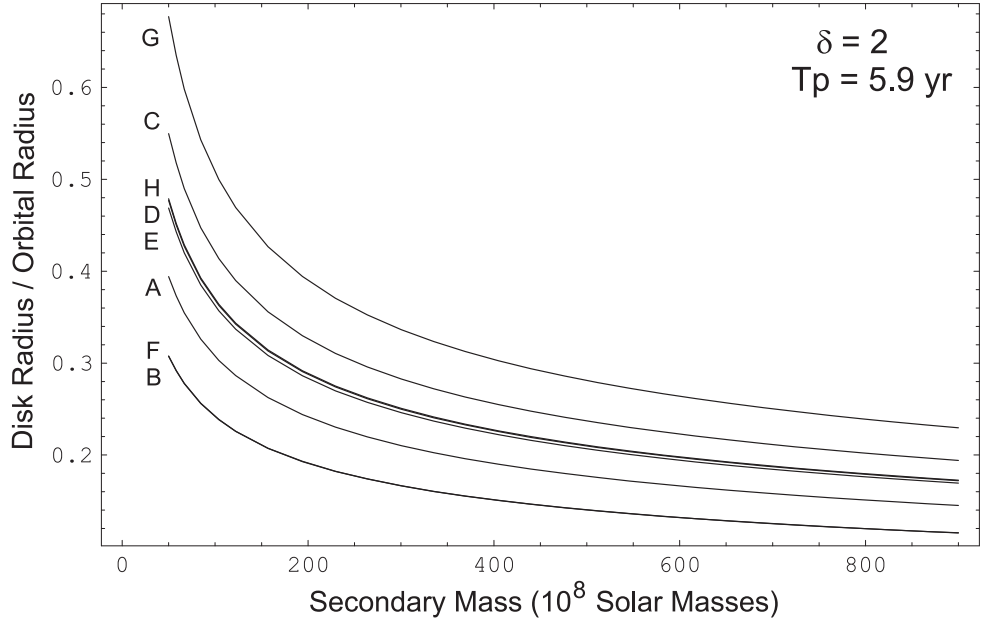} \caption{
Ratio of the precessing disk to orbital radius as a function of
the secondary black hole mass for the different primary models
defined in Table~1. The jet Doppler factor is 2 and the precessing
period in the source frame is 5.9 yr. }  \label{f2}
\end{figure}

\begin{figure}
\vbox to7.2in{\rule{0pt}{7.2in}} \includegraphics{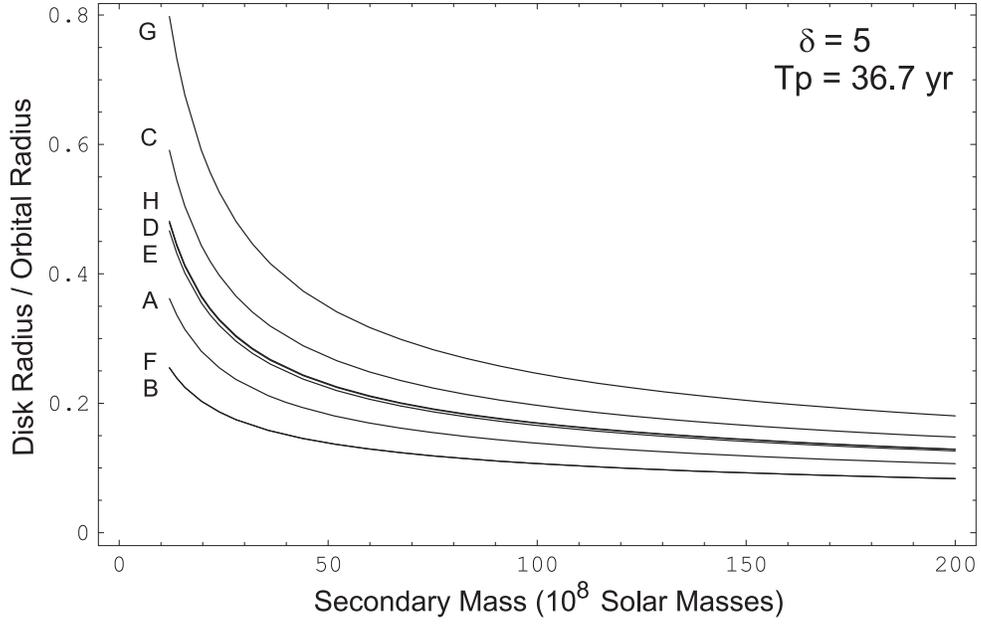} \caption{
Idem Fig. \ref{f2} but for a jet Doppler factor of 5. The
corresponding precessing period in the source frame is 36.7 yr. }
\label{f3}
\end{figure}

\begin{figure}
\vbox to7.2in{\rule{0pt}{7.2in}} \includegraphics{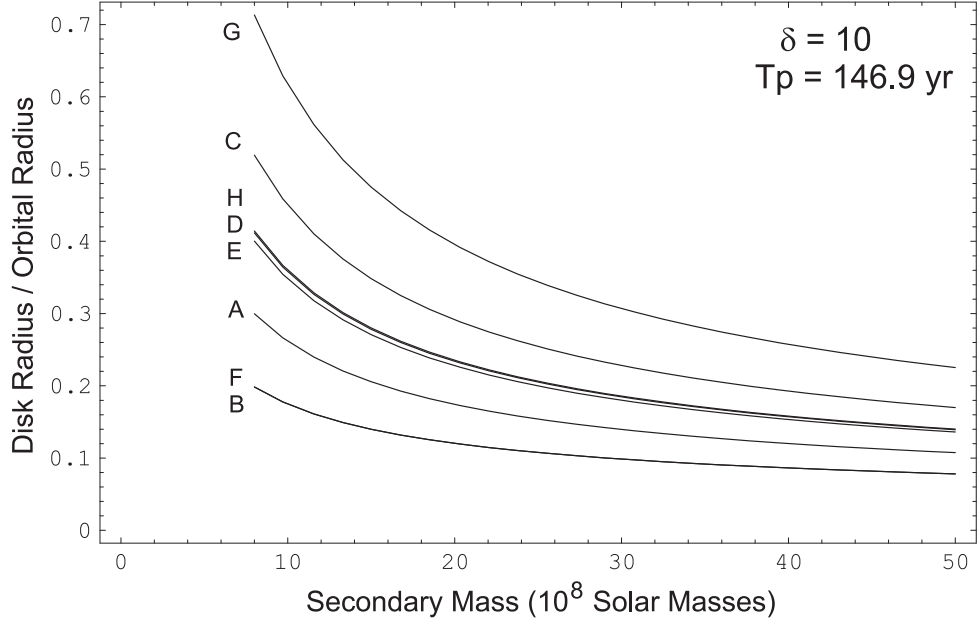} \caption{
Idem Fig. \ref{f2} but for a jet Doppler factor 10. The
corresponding precessing period in the source frame is 146.9 yr.}
\label{f4}
\end{figure}

The radius $r_{\rm d}$ of the precessing part of the disk is related to the disk angular
velocity $\omega_{\rm d}$ by $\omega_{\rm d}=(GM/r_{\rm d}^3)^{1/2}$. Then, using Eq.
(\ref{prec}) with the observational fact that $\cos\theta\sim 1$ (Fujisawa et al. 1999),
we can establish the ratio $\kappa=r_{\rm d}/r_{\rm m}$ for each model. Only models for
which $\kappa$ is significantly smaller than 1 can display uniform, nearly rigid,
precession in their inner accretion disks (e.g. Katz 1973, Larwood 1998). Since we have
only an upper bound onto the mass of the secondary black hole, we have plotted the curves
$\kappa=\kappa(m)$ for all models whose primary masses are listed in Tables \ref{M1} --
\ref{M3}. These curves are presented as Figures \ref{f2} -- \ref{f4}. The value of
$\kappa$ corresponding to the largest possible value of $r_{\rm m}$ is given in the last
column of Tables \ref{M1} -- \ref{M3}.

From Figures \ref{f2} to \ref{f4} we see that models with single
temperature disks and large outer radii (models F and B) are more prone to display
uniform precession. Models with $\delta=10$ in general imply $M>m$. Nonetheless, some models with $\delta= 2$ and $5$ are possible with $m>M$. We cannot exclude these models based only on a priori considerations. It is possible to imagine, for instance, that the original merger that resulted in the formation of the binary system occurred between an active blazar and a ``dormant" quasar with a very massive central object that already swallowed up all available gas. Deep observations of the host galaxy can shed some light onto this particular point. The detection of peculiar Fe K$\alpha$ line profiles can be used to test whether the secondary has or not associated an accretion disk (Yu \& Lu 2001).

The sound speed in the disk of the primary can be approximated by $c_{\rm s}\sim H
\omega_{\rm d}$, with $H$ the disk half-thickness. In the inner
disk, the constraint of efficient physical communication in the fluid imposes $c_{\rm s}>
\Omega_{\rm p} r_{\rm d}$. Then, we have that in this region:

\be H>\frac{\Omega_{\rm p}}{\sqrt{GM}}\;r_{\rm d}^{5/2}.  \ee

In our models for AO 0235+16 we found that $H\geq 5\times 10^{12}$ cm.

\section{Additional comments}

Precessing jet models are not the only kind of models based on
binary black hole systems that can explain periodic behavior in
AGNs. Other alternatives are orbital motion of a secondary black
hole with an associated jet (e.g. Rierger \& Mannhaim 2000, De
Paolis et al. 2002), accretion disk instabilities (Fan et al. 2001
and reference therein ), and pair jets (Villata et al. 1998). The
main difference between these models and the precessing jet model
is that the latter implies the precession of the innermost part of
the accretion disk.

Compton reflection of hard X-ray emission in the
cold, outer material can be expected in these kind of systems. The
iron K$\alpha$ line should change in the observer frame,
oscillating around an average value with a period $\sim
\Omega_{\rm p}$. Periodic changes in the intensity of the line
with amplitudes up to $\sim 30$ \% should be also observed (Torres
et al. 2003). Although the detection of the iron
K$\alpha$ line in AO 0235+16 is beyond the current sensitivity of
X-ray observatories, future space missions like Constellation-X,
with its superb spectral resolution and sensitivity might detect
the line and its oscillation, then providing a tool to probe the
nature of the periodic phenomena in this blazar, and the
particular kind of models discussed here.

\section{Conclusions}

We have analyzed the feasibility of precessing disk models for a binary black hole system
in the extremely variable BL Lacertae object AO 0235+16. We have presented an improved
determination of the black hole mass of the accreting object in the system. We have also
determined values and bounds to all other relevant parameters in the system. We found
that, if the periodic activity recently reported for this object is based on jet
precession induced by the gravitational torque of the companion black hole on the disk,
then a large variety of models are possible, including models where either the mass of the
primary is larger than the mass of the secondary or vice versa. If the Doppler factor of
the relativistic flow is $\sim 10$, as suggested by some authors, then $M>m$ for almost
all possible models. We emphasize, however, that the ultimate nature of the central engine
in this violent blazar remains an open problem. The models here proposed can be tested
through some simple predictions for the periodic behavior of the Fe K$\alpha$ line.
Hopefully, future observational constraints from long-term multifrequency monitoring and
space X-ray observations will help to unveil the source of the periodic events reported
for AO 0235+16.

\begin{acknowledgements}

GER research on high energy astrophysics is mainly supported by
Fundaci\'on Antorchas. Additional support comes from the Argentine
agencies CONICET (PIP 0430/98) and ANPCT (PICT 98 No. 03-04881).
JHF thanks the financial support from the National Natural
Scientific Foundation of China (grant No. 19973001, 10125313) and
the National 973 project (NKBRSF G19990754).

\end{acknowledgements}


{}

\end{document}